\documentstyle[preprint,eqsecnum,aps]{revtex}
\begin{document}
\draft
\tighten
\preprint{UCSBTH-96-11, hep-th/9605224}
\title{Counting States of Black Strings with Traveling Waves}
\author{Gary T. Horowitz\cite{gary} and Donald Marolf\cite{don}}
\address{Physics Department, University of California,
Santa Barbara, California 93106} \date{May, 1996}
\maketitle

\begin{abstract}
We consider a family of solutions to string theory which depend
on arbitrary functions and contain regular event horizons.
They describe six dimensional extremal black strings with traveling waves
and have an inhomogeneous distribution of momentum along the string.
The structure of these solutions near the horizon is studied and the
horizon area computed. We also count the number
of BPS string states at weak coupling whose
macroscopic momentum distribution agrees with that of
the black string.
It is shown that the number of such states is given by
the Bekenstein-Hawking entropy of the black string with traveling waves.

\end{abstract}

\newcommand{\f}{\dot f}
\def \k {\kappa^2}
\newcommand{\al}{\alpha}
\newcommand{\p}{\partial}
\newcommand{\V}{{\sf V}}
\newcommand{\be}{\begin{equation}}
\newcommand{\ee}{\end{equation}}

\vfil
\eject

\section{Introduction}

Recent progress in string theory has led to a statistical explanation of
the Bekenstein-Hawking entropy for a certain class of black holes [1 - 8].
(For a recent review, see \cite{hor}.) This
class includes extremal and near extremal, 
four and five dimensional
solutions. The microstates of these black holes turn  out to be associated with 
fields moving on a circle in an
internal spacetime direction.  To see a closer
connection between the quantum states and the classical spacetimes, it is
convenient to rewrite the black hole solutions so as to
explicitly include
the internal direction. This results in 
{\it black string} solutions in one higher dimension. The simplest
possibility (and the only one to be considered so far in connection
with entropy calculations) is for the black string to be translationally
invariant along the extra dimension. 

In addition to the spacelike (and timelike) translational symmetry,
the extremal limit of these black string solutions has a full
$1 +1$ dimensional Poincare invariance, including in particular
a null translational symmetry.  It has been known for some time \cite{GV,Garf}
that one can add traveling waves to any such solution, i.e., waves moving
along the string. Since the initial 
wave profile
can be specified arbitrarily, one obtains a class of solutions
having event horizons and labeled by free functions. These solutions
are quite remarkable in that the waves propagate indefinitely without
radiating to infinity or falling into the horizon.  Furthermore, 
the waves carry a finite energy and momentum per unit length along the string.
We will show that 
that the Bekenstein-Hawking entropy of each such solution
can be reproduced by counting string states, provided only that the 
free functions do not vary too rapidly. 
This restriction turns out to be necessary for a meaningful comparison
of the two entropies.
Our results greatly increase the
class of solutions with horizons for which such a counting has
been performed.

Black strings with traveling waves have recently been discussed in
connection with a different approach to understanding black hole entropy.
In \cite{lawi} 
it was suggested that these waves should be viewed as ``classical
hair" and an attempt was made to count the number of states associated
with this hair. This approach was further developed in \cite{cvts,tse}. 
The idea was that if the extra
dimension was sufficiently small then this hair would not be directly
visible but the degrees of freedom associated with it might account for
the black hole entropy. 

Although we start with the same classical solutions,
our interpretation
is quite different. First of all, it was suggested in \cite{lawi}
that since the dependence of these traveling waves on the radial coordinate
is fixed, one should view them as degrees of freedom on the horizon.
But since the event horizon is a null
surface, there is no sense in which a wave propagates along the 
horizon. We will study the global structure of these 
solutions and show that the  horizon is always {\it homogeneous}.
There is no local geometric
quantity which distinguishes one point on the horizon from any other:
There is no ``classical hair" on the horizon. Secondly, in \cite{cvts,tse}
it was proposed that this classical hair could be counted by studying a
fundamental string
propagating in the background of a  black string without any waves. The
(left-moving) states of this string correspond to linearized waves on
the black string background.
In contrast, we will study waves of finite amplitude.
Finally, we will consider the regime
where the size of one internal direction is large. Then for a {\it fixed}
wave profile, the black string still has a horizon with finite area
(which depends on the wave).
The usual semiclassical arguments imply that it has a nonzero entropy 
which must still be explained.
We will see that this entropy is naturally understood using the same
methods that have been applied to the black string without traveling waves.

For simplicity we will consider a six dimensional black string carrying
electric and magnetic charges associated with the Ramond-Ramond (RR)
three form $H$. In the
limit of weak string coupling, RR charges are carried by D-branes 
\cite{pol,pcj}, and 
the black string is described by a bound state of
D-fivebranes and D-onebranes. We will
study two different types of traveling waves. One corresponds to a wave $p(u)$
which
just changes the momentum density along the string.
The case of constant $p$ corresponds to the 
translationally invariant solution obtained by boosting the nonextremal
black string in \cite{host}.
It has been shown that in the limit of weak string
coupling (and for a single fivebrane), the states of the extremal
black string are
obtained by taking
a certain set of fields on a circle (which, in simple cases,
can be viewed as representing
the oscillations of
the onebranes inside the fivebrane),
and  considering all right moving modes with the given
total momentum.
The number of such states agrees with the exponential
of the Bekenstein-Hawking entropy in the limit of large charges and momentum
\cite{stva,cama,host}.
To represent the states of the black string with a momentum wave,
one would expect to keep the same fields on the circle, but now to
consider only those states with momentum distribution given by $p(u)$.
We will show that the number of such states is indeed the exponential of
the Bekenstein-Hawking entropy.

The second type of wave we will consider consists of an oscillation
$f^i(u)$ in a
four dimensional compact internal space (so that, together with the
six dimensions of the black string, one forms a ten dimensional
spacetime).  When this type of oscillation is present, the asymptotic momentum
distribution depends on both $p(u)$ and $f^i(u)$. 
Unlike $p(u)$, we will show
that this wave does not affect the area of the event horizon. 
That is, although their momenta differ, a
black string with waves $(p,f)$ has the same horizon area as
a black string with waves $(p,0)$.

This can
also be understood from counting states at weak string coupling.
When $f^i(u) \ne 0$, the black string has nonzero components of the momentum
density in the four internal directions. In the weak coupling $1+1$ dimensional
description, this corresponds to the {\it field} momentum $\pi$
canonically conjugate to the field $\chi$ that describes
the fluctuations of the onebranes inside the fivebranes.
We must thus count states with a given
distribution of this field momentum, as well as a given distribution of
the momentum along the string.
While these internal waves increase the momentum distribution along
the string (potentially increasing the number of available states),
we will see that the constraint on the field momentum
is a strong restriction on the allowed states.
The net effect turns out to be that, to leading order
for large charges, the number of available states is unaffected by
this internal wave.

One can also turn this around. Suppose we start with the weakly
coupled D-brane description in which BPS states are described by right
moving modes in $1+1$ dimensions. One should be able to specify any
macroscopic property of these states and increase the string coupling
to form an event horizon. If one fixes the total momentum, one forms
the usual translationally invariant black string. If one fixes the 
longitudinal momentum distribution one forms the black string with 
traveling wave $p(u)$. If one fixes the field momentum, one forms
the black string with the internal wave $f^i(u)$. In this way, one can see
various properties of the black string microstates in the classical solutions.

Adding  traveling waves to the six dimensional extremal
black string is analogous to adding
angular momentum (in two orthogonal planes with $J_1 = -J_2$ \cite{bmpv}).
In both cases
one does not break supersymmetry. The quantum states  of the resulting
solutions are a subset of the BPS states of the original nonrotating 
black string without traveling waves. One just restricts to states with
 given momentum distributions in the first case, or a given angular
momentum in the second. However, while the angular momentum is
specified by a single number, the momentum distributions are  given
by arbitrary functions.

In the
next section we describe the black string with
traveling waves and study the structure of its event horizon;
certain technical details are relegated to
the appendix. In section three we  count the corresponding states 
in string theory at weak coupling,
and section four contains some discussion of the
results.

\section{The Black String Solutions}
\label{metric}

We start with the type IIB string theory in the ten dimensional Einstein
frame, keeping
only the metric, dilaton and RR three form $H$:
\be \label{action}
S = {1\over 16\pi G_{10}} \int d^{10} x \sqrt{-g}\left(
R -  {1\over 2} (\nabla \phi)^2 
- {1\over 12}  e^\phi H^2 \right )
\ee
where $G_{10} = 8\pi^6 g^2$ is the ten-dimensional
Newton's constant in units with $\alpha'=1$, $g$ denotes the
string coupling, and 
the zero mode of $\phi$ is defined so that $\phi \rightarrow 0$ 
asymptotically. 
We wish to consider toroidal compactification to five dimensions
with an $S^1$ of length $L$ and a $T^4$ of volume 
$\V =(2\pi)^4 V$.\footnote{We 
follow the conventions of \cite{hms} so that T-duality sends $V$ to $1/V$ and
S-duality sends $g$ to $1/g$.}
We will assume that $L >> V^{1/4}$ so that the solutions resemble
strings in six dimensions.  This will be of use in section
\ref{cs}.
We also impose spherical symmetry in the five noncompact dimensions.
The electric and magnetic charges associated
with $H$ are defined by
\be
Q_1 = {V\over 4\pi^2 g} \int e^\phi *H,    \qquad
Q_5 = {1\over 4\pi^2 g} \int H
\ee
and are normalized to be integers.
Below, we will describe a family of black string solutions with two
kinds of traveling waves and study the effect of each type of
wave in turn on the horizon geometry.

\subsection{Black Strings and Traveling Waves}

A six parameter family of solutions to (\ref{action}) with the
desired properties
and regular event horizons was found in
\cite{hms,cvyoum}. Here we will be interested in the extremal limit of these
solutions. We will also adjust one of the parameters so that the volume of the
four torus is constant.\footnote{In the notation of \cite{hms}, we set
$\alpha = \gamma$.}  The resulting solution is 
\be\label{simple}
ds_0^2 = \left(1 + {{r_0^2} \over r^2}\right)^{-1}
 [-du dv + {p\over r^2} du^2 ]
+  \left(1+ {{r_0^2} \over
r^2}\right)( dr^2 + r^2 d\Omega^2_3 ) +dy_i dy^i
\end{equation}
with
\be\label{charge}
Q_1 = {Vr_0^2\over g}, \qquad Q_5 = {r_0^2\over g}
\ee
and $\phi=0$. The coordinates $y^i$ label points on the $T^4$ and $u=t-z, \ 
v=t+z$ where $z$ is a 
coordinate on the $S^1$.

The black string (\ref{simple}) has an
event horizon at $r = 0$ which is a smooth surface (for
$p > 0$; see appendix) of  area $2\pi^2 r^2_0  
L\V \sqrt{p}$. The string carries an ADM energy and momentum
\be
 E = {L(2r_0^2 +p) \over \k}, \qquad P = {L p\over \k}
\ee
where
\be\label{defkap}
\k = {4G_{10} \over  \pi\V} = {2\pi g^2 \over  V}
\ee
Note that for
$p < -2r_0^2$ the string carries negative energy and for $p <0$
the compactified
spacetime has closed timelike curves outside the horizon.  In
addition, the compactification introduces a conical singularity when
$p=0$ \cite{ght}.  As a result,
we will consider only the case $p >0$.

Since the Killing vector field ${{\partial} / {\partial v}}$ is null,
traveling waves can be added to this metric using the method of 
Garfinkle and Vachaspati \cite{GV,Garf}.  The resulting solutions
have been studied in \cite{lawi,cvts}.
The metrics take 
the form
\begin{equation}
\label{inhom}
ds^2 = \left(1 + {{r_0^2} \over {r^2}}\right)^{-1} [-dudv + 
K(u,x,y) du^2] +\left(1 + {{r_0^2} \over {r^2}}\right) dx_i dx^i +dy_i dy^i
\end{equation}
where we have introduced the cartesian coordinates $x^i$ in the four
noncompact spatial dimensions, i.e. $r^2 = x_i x^i$  with the
index $i$ being raised and lowered by the flat metric $\delta_{ij}$.
Note that (\ref{inhom})
again has a null Killing field ${{\partial} /{\partial v}}$
whose integral curves are null geodesics.  The metric disturbance $K$
propagates along these curves.

The metric (\ref{inhom}) satisfies the field
equations with the same matter sources
as  (\ref{simple})
provided
the waves $K(u, x,y)$
satisfy the appropriate equation in the eight transverse
coordinates $(x,y)$:
\begin{equation}
\label{Laplace}
 (\partial_{x_i} \partial_{x^i}
+\left[ 1 + {{r_0^2} \over {r^2}}\right]
 \partial_{y_i} \partial_{y^i}) K =0.
\end{equation}
Note that the waves may have arbitrary $u$ dependence. 
$K(u,x,y)$ may include terms with `sources' at $r=0$, since
this is a coordinate singularity in (\ref{inhom}). For monopole fields,
we will see that 
the horizon remains well defined, with no
actual sources present..

Many solutions to (\ref{Laplace}) simply correspond to gravitational
waves superposed on the black string. This can be seen from the fact that
they change little in the
limit $r_0 \rightarrow 0$ in which the black  string disappears.  Such waves
are not `anchored' to the string in any way.  Other waves will
not be of interest as the resulting spacetimes do not
have five asymptotically flat directions.  However, there is a
class of solutions (studied
in \cite{cmp} and  \cite{DGHW} for the fundamental string)
for which the metric remains asymptotically flat and the 
waves {\it are} anchored to the string.
Keeping spherical symmetry in the
asymptotically flat directions,
this is the class for which
\begin{equation}
\label{waves}
K(u,x,y) = {{p(u) } \over {r^2}} - 2 \ddot{f}_i(u) y^i.
\end{equation}
One could also consider the solution $K= -2\ddot{h}_i(u) x^i$. These waves
will be briefly discussed in 
section \ref{disc} and in the appendix.
The dots $(\ \dot{}\ )$ above represent derivatives with respect to $u$, and 
writing the waves in this way will simplify the notation below.

The waves described by $p(u)$ and $f_i(u)$ are rather different, both in
their effect on the metric and in the way we will interpret them in terms of
BPS states.
As such, it is useful to give them different names.
Those described by $p(u)$ will be called `longitudinal waves,' as
they carry only momentum directed along the large $S^1$.  
Those described by $f_i(u)$ will be called `internal
waves' as they carry momentum components in the internal directions.

With $K$ given by (\ref{waves}) the metric appears neither asymptotically
flat, nor translationally invariant in the internal four-torus. Both of
these difficulties can be resolved by introducing new coordinates
$(u,v',x,y')$ which are related to those
above through
\begin{eqnarray}
\label{afct}
v' &=& v + 2 \dot{f}_iy^i + \int^u \dot{f}^2 du \cr
y'{}^i &=& y^i + f^i
\end{eqnarray}
where $\dot{f}^2 = \dot{f}_i \dot{f}^i$.  The metric then takes
the form
\begin{eqnarray}
\label{af}
ds^2 &=& \left(1 + {{r_0^2} \over {r^2}}\right)^{-1} [-dudv' + 
{{p(u) + r_0^2 \dot{f}^2} \over {r^2}} du^2]
+  \left(1+ {{r_0^2} \over
{r^2}}\right)(dr^2 + r^2 d\Omega^2_3) \cr
&-&  {2{r_0^2}
\over {r_0^2 + r^2}} \dot{f}_idy'{}^idu  + dy'{}_idy'{}^i.
\end{eqnarray}
It is in terms of these coordinates that we make the periodic
identifications.  The large $S^1$ is defined
by the identification $z\rightarrow z - L$, or $(u,v',x,y') \rightarrow
(u + L, v' - L, 
x,y')$, and the small four-torus is defined by the identifications
$(u,v',x,y') \rightarrow (u,v',x,y' + a_I)$
for an appropriate set of vectors $a_I$, $I=1,2,3,4$.
Clearly, $p(u)$ and $\dot{f}_i(u)$ must be periodic with period $L$.
In addition, we will take $f_i$ itself to be periodic.  In the 
ten dimensional space before compactification, this 
amounts to choosing coordinates in which the string has no net momentum
in the internal directions.

The asymptotic charges can be read directly from the metric (\ref{af}).
The black string carries a longitudinal momentum
\be
P = {1\over \k} \int_0^L  (p + r_0^2 \dot{f}^2) du
\ee
where $\k$ is given by (\ref{defkap}).
So $(p + r_0^2 \dot{f}^2) /\k$ is naturally
interpreted as a momentum density along the string.  Similarly, 
although the total momentum vanishes in the the internal ($y$) directions,
the oscillations give the string a nontrivial effective internal momentum
distribution $r_0^2 \dot{f}_i/\k$.   Note that, in
a weak coupling limit where localized energy-momentum is well defined,
a string-shaped matter source with energy 
density $(2r_0^2 + p + r_0^2 \dot{f}^2 ) /{\k}$,
momentum density $(p +
r_0^2 \dot{f}^2) / {\k}$ in the string direction, and momentum
density $(r_0^2 \dot{f}_i)/ {\k}$ in the internal directions
will produce a gravitational field with the same asymptotic behavior as
(\ref{inhom}).  As a result, (\ref{inhom}) is the black string one would 
expect to obtain by taking a collection of D-strings and D-fivebranes
with a fixed profile for the momentum components 
and turning up the coupling.  

There is an important
distinction between the black string with traveling waves described above
and a fundamental string with traveling waves. It was shown in \cite{DGHW}
that in order for the traveling wave (\ref{waves}) to match onto a 
fundamental string source, one must set $p(u) =0$. The source
string then carries a wave with profile $f^i(u)$ (note that this fixes the
normalization of $f$). Physically, one should expect $p(u)$ to vanish 
since it represents the longitudinal momentum, and a 
fundamental string has only transverse degrees of freedom.
More precisely, $p(u)$
is the eigenvalue of the worldsheet stress energy operator $T_{uu}$
which vanishes by reparametrization invariance. In \cite{lawi}, it was
suggested that there is an analog of the corresponding
level matching condition
$L_0 - \bar L_0 =0$ even for the black string. We find no such 
restriction here. We will see that the horizon is well defined for
all waves (\ref{waves}).

\subsection{Longitudinal Waves Near the Horizon}
\label{long}

We begin our study of the horizon by setting $f_i = 0$ and
considering the case of purely longitudinal waves.  The effects of
internal waves will be considered in section \ref{int}. The metric
is thus (\ref{inhom}) with $K= p(u)/r^2$.
We wish in particular to compute the horizon area for comparison
with a counting of weakly coupled string states. One might be
tempted to compute the horizon 
area by setting $t$ equal to a constant
(where $u=t-z,\ v=t+z$) and $r=0$, to obtain the  area 
$2\pi^2 r^2_0 \V\int_0^L \sqrt{p}$; however we will see that this
is in general not the correct result.
The above reasoning fails due to the fact that the constant $t$ surfaces
do not intersect the horizon. However, the exact expression will
reduce to this result in an interesting limit.

For simplicity, we work here with only the leading order 
behavior of  
(\ref{inhom}) near the horizon;  a more complete analysis will be given
in the appendix.  While understanding the leading order behavior is not
always sufficient, it does give the correct horizon area in this case
and is much simpler than the full treatment.
Near the horizon, the metric takes the form
\begin{equation}
\label{hor}
ds^2 = r_0^2 \bigg\{ R^{-2} (-dudv + dR^2) + {{p(u)}\over {r_0^4}}du^2 + 
d\Omega^2_3 \biggr\} + dy_i dy^i
\end{equation}
where $R = r_0^2/r$  and
we have neglected subleading terms in $1/R$.
It is useful to 
realize that the (future) horizon lies not only at $R = \infty$, but also at 
$u,v = \infty$.  This can be seen from  the fact that the horizon
should lie at $t= \infty$ and $u =  t- z$ while $v = t+z$.  More
precisely,
it may be shown that all geodesics (timelike, null, {\it and}
spacelike) which reach $R= \infty$ do so only as $u$ and $v$
diverge.

To study the horizon, let us
suppose that there exist functions $\sigma(u)$, $G(u)$, and $F(u)$ satisfying
$r_0^{-4}p = \sigma^2 + \dot{\sigma}$, 
$\sigma = {\partial}_u \log G$, 
and $G = (\dot{F})^{-1/2}$, where again the dot ($\ \dot{} \ $)
denotes $\partial_u$.
Introducing new coordinates $U = F(u)$, $V = v - \sigma R^2$, and 
$W = G(u)/R$, the metric
(\ref{hor}) takes the form
\begin{equation}
\label{h2}
ds^2 = r_0^2 \biggl\{ - W^2 dU dV + W^{-2} dW^2 + d \Omega^2_3 \biggr\}
+ dy_i dy^i
\end{equation}
which is just three dimensional anti-deSitter space cross
$S^3\times T^4$.  Note that the local geometry is completely independent of the
wave profile\footnote{Metrics with different longitudinal
waves $p_1 \neq
p_2$ are not diffeomorphic, but the difference appears only in the 
departure from the asymptotic form  (\ref{hor}).} $p(u)$.

In order for expression (\ref{h2}) to be valid, we must show 
that the functions  $\sigma$, $G$, and $F$ in
fact exist.  Consider first the equation $ r_0^{-4} p(u) = \sigma^2 + 
\dot{\sigma}$.
Recall that we consider periodic $p$ with $p \ge 0$. 
Suppose that we try to solve this equation subject to the boundary condition
$\sigma(0) = 0$.  As long as $\sigma^2 < r_0^{-4} p$, we see that 
$\dot{\sigma} >0$ and $\sigma$ is increasing.  Similarly, 
when $\sigma^2 > r_0^{-4} p$, we have $\dot{\sigma} < 0$.
Large departures of $\sigma$ from $r_0^{-4} p$ decay as $1/u$.  
As a result, the function
$\sigma$ simply follows $r_0^{-2}\sqrt{p}$ and stays within a bounded region
as $u$ ranges over the positive real line.
A smooth solution $\sigma(u)$ exists for all $u$ and it is reasonable to assume
that this solution asymptotically approaches a periodic function
$\sigma_0(u) \ge 0$ which also solves the differential equation.  From
now on we will work with the periodic solution $\sigma(u) = \sigma_0(u)$.
Independent of this assumption, one can always start with any periodic
$\sigma(u)$, and consider the wave given by $p(u)=r_0^4(\sigma^2+\dot{\sigma})$.

Since $\sigma \ge 0$ is continuous and $G = \exp{\int \sigma \ du}$, $G$ also 
exists and is positive for all $u$.  In addition, $G$ diverges exponentially as
$u \rightarrow \infty$.  It follows that $F = \int G^{-2} du$ in fact
{\it converges} so that the horizon is located at a finite value 
of $U$ which  we will take to be $U = 0$.
The set $(V,w = r_0\log W,y)$ together with the angles $\theta$ on the
3-sphere define good coordinates on this surface
and the metric on the horizon is just
\be
ds^2_{U = 0} = dw^2 + 
r_0^2 d\Omega^2_3 + dy_i dy^i.  
\ee

The area of the horizon depends on the range of $w$ which is determined
as follows.
The periodic identification $ (u,v,x,y) \rightarrow (u +L,v-L,x,y)$ 
of the large $S^1$ induces the
identifications $(u,v,R,\theta, y) \rightarrow (u + L, v-L,R,\theta,y)$ and
$(U,V,w,\theta,y) \rightarrow (U + L{\overline{G^{-2}}}(u), V - L,
w + r_0L \overline{\sigma},\theta,y)$ where
$\overline \sigma = L^{-1} \int_u^{u+L} \sigma(u')du'$
and $\overline {G^{-2}}(u) = L^{-1}  \int_u^{u+L} G^{-2}(u')du'$.  
Since $\sigma(u)$ is periodic, $\overline{\sigma}$ is
independent of $u$. Thus $w$ has period $r_0 L \overline{\sigma}$, and 
the horizon of (\ref{hor})
is a smooth surface with area 
\begin{equation}
A = 2\pi^2 r_0^4 L\V \overline{\sigma}.
\end{equation}

Note that since $\sigma$ is periodic the total momentum is $P = \kappa^{-2}
r_0^4 \int_0^L \sigma^2 du$.  Also, by the Schwarz inequality we
have
\be
L \overline{\sigma} = \int_0^L {\sigma} du \le  \left ( L \int_0^L \sigma^2
du \right)^{1/2}
\ee
with equality when $\sigma$ is constant.  This means that
for a fixed total momentum $P$ the horizon area
is maximized by the uniform distribution $\sigma = r_0^{-2} \sqrt{p}
= constant$.

It is of interest to consider a limit in which $p$ is, in some
sense, slowly varying.  Note
that when $\sigma^2 \gg \dot{\sigma}$, we have $\sigma
 = r_0^{-2} \sqrt{p}$ to leading order.
Thus, the proper condition for $p$ to vary slowly is 
\be\label{slow}
p^{3/2} \gg r_0^2 |\dot{p}|. 
\ee
In this case, the  area of the
horizon is given by the naive result $A = 2\pi^2 r_0^2\V
\int_0^L {\sqrt{p}} du$, and the Bekenstein-Hawking entropy is
\be\label{entropy}
S_{BH} = {A \over 4 G_{10}} = \sqrt{2\pi Q_1 Q_5}\int_0^L {\sqrt{p/\k}} du.
\ee
In particular, if $p$ is constant, $p/\k=P/L \equiv 2\pi N /L^2$, the 
Bekenstein-Hawking entropy reduces to the familiar form 
$S_{BH} = 2\pi\sqrt{Q_1 Q_5 N}$.
We will mostly consider the case when (\ref{slow}) is satisfied.

Note that the horizon (at $u = \infty$) for the metric (\ref{hor})
is a smooth surface despite the fact that the periodic
profile function $p$ has
no well-defined value for large $u$.  This is because, in the absence
of identifications, the
function $p$ is pure gauge in the spacetime (\ref{hor}). It enters only
in determining the periodic identification along the black string.
For the full metric (\ref{inhom}), the function $p$ is again locally
pure gauge near the horizon even though it carries physical information
(e.g. the momentum distribution) far from the horizon.
It is shown in the appendix that this spacetime is at least 
$C^0$ near the horizon and that the local horizon geometry is
independent of $p$.  

As a final comment, note that since
$U = F(u)$ is an outgoing null coordinate and the horizon lies at
$U = 0$, 
the disturbances near the horizon are not
traveling {\it along} the horizon but
are in fact following {\it outward} directed null geodesics.
That these disturbances appear to be traveling along the horizon 
when described through the metric (\ref{inhom}) is a result of the
fact that angles are not Lorentz invariant and that natural coordinates
near infinity differ from natural coordinates near the horizon by a 
divergent Lorentz boost.  This results in an extreme version of the
`headlight effect' (familiar from radiation produced by a rapidly moving
charge) in which radiation distributed across a large angle 
in one frame is in fact distributed across a small angle  in another.

\subsection{Internal Waves Near the Horizon}
\label{int}

It is not difficult to extend this discussion to include internal 
waves $f_i$ as well.  Since the term involving $p(u)$ clearly dominates the one
involving $f_i$ in (\ref{waves}) near the horizon $r=0$, one might expect that
the structure of the horizon is independent of the internal waves. We will
see that this is indeed the case.

After shifting the origin of $v$ in (\ref{inhom})
by $2\dot{f}_iy^i$, we see that
adding the internal waves corresponds to simply adding the term
$2(1 + {{r_0^2} \over {r^2}})^{-1} \dot{f}_idy^idu$ to the
leading order metric
(\ref{hor}) near the horizon; again, we take $f_i(u)$ to be
periodic.  In terms of the coordinates $(U,V,W,\theta,y)$
this is just
\begin{equation}
\label{hto}
2r_0^2 W^2 \dot{f}_i dy^i dU
\end{equation}
to leading order in $1/R$.
The term (\ref{hto}) is not smooth on the horizon since
the periodic function
$\dot{f}_i$ is not well-defined at $U=0$ ($u=\infty$).
However, the metric can be written in $C^0$ form
(which is smooth enough for our purposes) by a
further transformation to coordinates $(U,\tilde{V},W,\theta,\tilde{y})$
where
\begin{eqnarray}
\tilde{V} &=& V - W^2 r_0^2 \int^U \dot{f}^2 dU \cr 
\tilde{y}^i &=& y^i + r_0^2 W^2 \int^U \dot{f}^i dU. 
\end{eqnarray} 
Note that $\int^U \dot{f}_i dU = \int^{u} \dot{f}_i  G^{-2}
du$ converges
as $u \rightarrow \infty$ ($U \rightarrow 0$) and so is
$C^0$ at the horizon.  The same is true
of $\int^U \dot{f}^2 dU$.  The actual form of this metric is not
particularly enlightening.

In achieving the $C^0$ form it was not necessary to replace the
coordinate $U$.  As a result, the horizon still lies at $U=0$.
Furthermore, since (\ref{hto}) is proportional to $dU$,
it does not change the local geometry of any $U = constant$ surface 
where it is well defined.  Since the
exact metric is $C^0$ in $U$ at the horizon $U=0$, it follows that the
metric on the horizon is again given by
\begin{equation}
 ds^2 = dw^2 + r_0^2 d\Omega^2_3 + dy_i dy^i
\end{equation}
and, in particular, it is still homogeneous.  As before, this is due
to the fact the that waves do not propagate {\it along} the the horizon, 
but instead become purely outgoing near $U=0$.

In addition, because the functions $\dot{f}_i$ are periodic, it may
be verified that the internal waves effect the coordinate
identifications 
on the horizon only through the coordinate $V$; the
identifications of the coordinates $(w, \theta,y)$ remain
unchanged.  However, 
${{\partial }\over {\partial V}}$ is the null generator of the horizon
and $V$ plays no role in determining the area.  As a result, the 
horizon area is completely 
independent of the internal wave profile 
$f_i$.

\section{Counting states}
\label{cs}

Let us first recall the counting of string states for the black string
without traveling waves i.e. constant $p$ 
and $f_i =0$. The idea is to go to weak string coupling where the
RR charges are carried by D-branes. One starts with a ten dimensional
flat spacetime with five
directions compactified on a torus for which one circle of length $L$ 
is much larger than than rest. The black string of the previous section
corresponds to  bound states of
$Q_5$ fivebranes and
$Q_1 $ onebranes wrapped around the torus.
Since one compact direction is much larger than the
rest, the excitations of these D-branes are described by a $1+1$
dimensional sigma model. 
Using the usual D-brane technology,
one finds  $4Q_1 Q_5$ bosonic
fields and an equal number of fermionic fields on the circle. (For the
case $Q_5 =1$, one can view this as arising from the oscillations of the
$Q_1$ strings in the four internal directions.) The number of
BPS states with total momentum $P = 2\pi N/L$ is then $e^S$ where
$S = 2\pi \sqrt{Q_1 Q_5 N}$
in agreement with the Bekenstein-Hawking entropy for the case of no
traveling waves.

\subsection{Longitudinal Waves}
\label{clw}

The weak coupling limit of the black string with longitudinal waves
corresponds to states where the momentum {\it distribution} is fixed, not just
the total momentum.  Quantum mechanically, however, one cannot fix the
{\it exact} density $p(u)$ of momentum in a 1+1 field theory.
The reason is simply that $p(u)$ and $p(u')$ do not commute;
the Fourier modes of $p$ satisfy the Virasoro algebra.  We therefore 
take a `mesoscopic' viewpoint of our fields.  That is, we will imagine
we use apparatus which can resolve the system only down to a
`mesoscopic' length scale $l$ which is much larger than the
`microscopic' length scale (discussed below) on which quantum effects
are relevant.  We will therefore divide the spacetime into
$M = L/l$ intervals $\Delta_a$ ($a \in \{1,...,M\}$) of length
$l \ll L$.  If our instruments find a momentum distribution $p(u)$, 
this means simply that the total momentum in the interval $\Delta_a$
is $P_a =\kappa^{-2} \int_{\Delta_a}p(u) du$; we cannot resolve $p(u)$ on
smaller scales.  Of course, it would be meaningless for us to
assign a distribution $p(u)$ which has structure on scales of 
size $l$ or smaller.  As a result, $l$ should be much smaller
than $p/|\dot{p}|$, the `macroscopic' length scale set by the
variation of the wave profile $p$.

We can expect a state counting argument to yield the Bekenstein-Hawking
entropy only when such a mesoscopic length scale actually exists.
This will restrict us to a certain class of momentum distributions
$p(u)$.  We use the following heuristic argument to motivate the
appropriate condition on $p$; it will be shown below that the
resulting condition is in fact sufficient for the existence of our
mesoscopic picture.

If quantum correlations between the intervals are to be
irrelevant, 
the wavelength $\lambda$ of a typical excited mode should be much
less than $l$.
In the interval $\Delta_a$, each field carries a 
momentum  of order $pl/Q_1Q_5\k$ so we expect
$\lambda \sim {{Q_1Q_5\k} / {pl}}$.
As a result, 
we require
\begin{equation}
\label{meso}
p/|\dot{p}| \gg l \gg \sqrt{{Q_1Q_5\k} \over p }.
\end{equation}
Such an $l$ can exist only when
\begin{equation}
\label{pc}
p^{3/2} \gg |\dot{p}| \sqrt{Q_1Q_5\k} = {\sqrt{2\pi}} r_0^2
|\dot{p}|,
\end{equation}
so we consider only momentum distributions such that (\ref{pc}) holds.
This is just the condition (\ref{slow}) that $p$ be `slowly varying',
under which the Bekenstein-Hawking entropy is given by (\ref{entropy}).
This argument is heuristic since it does not rule out the possibility
that the momentum arises from a large number of quanta each with momentum
of order $1/l$; we will show below that this does not occur.
On the other hand, the above argument does show that
whenever $p$ satisfies (\ref{slow}), one can choose $l$ so that
$p$ is approximately constant on $l$ and
the level number $N \sim pl^2/Q_1Q_5\kappa^2$ of a typical
field is large in each interval.

If one could view states on the circle as consisting of a collection 
of $M$ independent systems of length $l$,
then
the number of states with momentum $P_a = 2\pi N_a/l$ on the 
$a^{th}$
interval would be $e^{S_a}$ with $S_a = 2\pi \sqrt{Q_1 Q_5 N_a}$, since
$N_a \gg Q_1Q_5$ by (\ref{meso}). The
entropy of independent systems is additive, so the 
total entropy for the system with momentum distribution 
$P_1, \cdots P_M$  would be
\be\label{dbrane}
S= 2\pi \sqrt{Q_1 Q_5} \sum_{a=1}^M \sqrt{N_a}. 
\ee
But $\sum \sqrt{2\pi N_a} = \int_0^L \sqrt {p/\k}$, so this agrees with the
Bekenstein-Hawking entropy of the black string (\ref{entropy}).
We now justify our
assumption that each interval acts like an independent system when
$l$ satisfies (\ref{meso}).

To begin,
consider a single scalar field $\chi$ on a circle of length $L$
parameterized by a (real) coordinate $z$. 
We want to count states with momentum
distribution $p(z)$  satisfying (\ref{slow}).
Let $P= \kappa^{-2}\int_0^L p$ be the total momentum and
let $\Delta_a$ be the
intervals introduced above. 
We now introduce a series of smooth cut-off functions
$g_a(z)$ which will project
operators into each segment of our circle. These functions satisfy
the following two properties:
(1) $g_a$ has support only in $\Delta_a$ and (2) $g_a= 1$ except for
a distance $\epsilon$ from each endpoint where $\epsilon < 1/P$. 
The bound on $\epsilon$ comes
from the fact that any state with support in the region where $g_a$ is
not constant has momentum larger than the total momentum we are considering.
Hence it will not contribute.    

The tension of the D-strings is $1/g$ times the usual string tension
($1/2\pi$ in units with $\alpha' = 1$),
so the action for $\chi$ is
\be
S = -{1\over 4\pi g}\int (\p \chi)^2.
\ee
Since the field on the entire circle has independent left and right moving
modes, it follows that in each segment there will also be independent left 
and right moving modes.  Because we wish to count BPS
states, we
will consider only, say, the right moving modes, so $\chi = \chi(u)$ with 
$ u = t-z$.
The local momentum density is given by
the stress energy tensor $T_{uu} = (\p_u \chi)^2/2\pi g $. The total
momentum in segment $\Delta_a$ is thus given by 
\be
P_a = \int_{\Delta_a} dz \ g_a \ T_{uu}.
\ee
We also define modes localized in each segment by
\be 
\alpha_{a,n} = {1\over \pi \sqrt{2 g}}\int_{\Delta_a} dz\ g_a\ e^{i2\pi nu/l}
\ \p_u \chi,
\ee
so that the field on $\Delta_a$ can be represented in the form
\begin{equation}
\label{fe}
\partial_u \chi = {\pi \sqrt{2 g} \over l} \sum_{n=-\infty}^{\infty}
\alpha_{a,n} e^{-i2 \pi n
u/l} + {\cal O}(\epsilon).
\end{equation}
Using the equal time commutation relations $[\p_u \chi(z) , 
\p_u \chi(\tilde z) ]= (i\pi g) \delta'(\tilde z -z)$,
one can show that $[\alpha_{a,m}, \alpha_{b,n}] = 
m \delta_{a,b} \delta_{-m,n} +{\cal O}(\epsilon)$.
In other words,  
the modes in different intervals commute, and within  each
interval they satisfy the usual commutation relations up to small correction
terms.  

One can 
define a vacuum state in each interval $\Delta_a$
by requiring that it be annihilated
by all modes $\alpha_{a,n}$ with positive mode number $n$.
Together with (\ref{fe}),
the commutation relations imply that 
for negative $n$ (and to leading order in  $\epsilon$),
the operators
$\alpha_{a,n}$ create states localized in region $\Delta_a$
with momentum $2\pi n /l$. It also follows that 
\be
P_a = {2\pi\over l} \left( {\alpha_{a,0}^2\over 2}  + \sum_{n=1}^\infty
\alpha_{a,-n} \alpha_{a,n}\right) + {\cal O}(\epsilon).
\ee
This is just the same as 
the relation satisfied by the total momentum and the usual modes
defined on the entire circle. (Recall 
that for the entire string $P$ is the Virasoro operator $L_0$.)
It then follows that 
the number of states in the
interval $\Delta_a$ with momentum $P_a= 2\pi N_a/l$ is given by $e^{S_a}$ where 
$S_a = 2\pi \sqrt{N_a/6}$ in the limit of large $N_a$.

The above argument counts the number of states relative to the vacuum 
state in each interval. To complete the counting we need to relate
these local vacua to the usual global vacuum state on the circle.  Note
that the local modes $\alpha_{a,n}$ are related to the
global modes $\alpha_n$ (defined on the entire circle of 
length $L$) through
\begin{equation}
\label{mr}
\alpha_{a,n} = {1\over L}\sum_m  \alpha_m
\int_0^L g_a e^{i2\pi nu/l} e^{-i2\pi mu/L}.
\end{equation}
The vacuum defined on the interval $\Delta_a$ is the state
annihilated by $\alpha_{a,n}$ for $n>0$ while the actual
vacuum of the system is annihilated by $\alpha_n$ for $n>0$. 
It can be shown from (\ref{mr}) that
the contribution of the creation operators $\alpha_m, \ m<0$
to the annihilation operators $\alpha_{a,n}, \ n>0$ is ${\cal O}(1/n)$,
so the vacua are essentially equivalent for the high modes.
We can use the 
equipartition theorem to estimate the
typical mode number $n$:  With overwhelming likelihood, the momentum
in mode $n$ should be roughly independent of $n$ up to some cutoff
$n= \Lambda_a$.  Suppose that this momentum per mode is ${{2\pi} \over
l}\tilde{N}_a$. Then for $n < \Lambda_a$ we have $nN_{n,a} = \tilde{N}_a$,
where $N_{n,a}$ is the occupation number of the $n^{th}$ mode
of the $a^{th}$ interval.  
Since the cutoff is set by the condition that 
the energy per mode ${{2\pi \tilde{N}_a} \over l}$ is
insufficient to excite the next
higher mode, we must have $\Lambda_a \sim \tilde{N}_a$.  It
follows
that the total level number satisfies $N_a \sim
\tilde{N}_a^2$, and that a typical excited mode has $n \sim 
\sqrt{N_a}$.
Thus the vacuum for these local modes agrees with the actual
vacuum to order $N_{a}^{-1/2}$.

Combining these results, we see that for a single scalar field
the log of the number of states with momentum 
distribution $p(u)$ is given by  $S= 2\pi \sum \sqrt {N_a/6}$. For the
weak coupling limit of the black string with traveling waves, one has
$4 Q_1 Q_5$ bosonic fields and an equal number of fermionic fields. The
entropy is thus given 
by\footnote{This is also a consequence of the equipartition
theorem, together with the fact that each fermion contributes as half a boson.
If we had $6 Q_1 Q_5$ independent bosonic fields with total
level number $N$, then it would be overwhlemingly likely that each field
has a level number $N/(6 Q_1 Q_5)$ and a corresponding
entropy $S_{single \ field} = 2 \pi \sqrt{{N} \over {36 Q_1 Q_5}}$. 
The total entropy is thus $S = 2 \pi \sqrt{Q_1 Q_5 N}$.} 
\be
S= 2\pi \sqrt{Q_1 Q_5} \sum_{a=1}^M \sqrt{N_a} 
= \sqrt {2\pi Q_1 Q_5}\int_0^L \sqrt {p / \k}
\ee
in agreement with the Bekenstein-Hawking entropy (\ref{entropy}).

Notice that since we have only considered right moving modes, these
are all BPS states. The number of such states is independent of the string
coupling, so we expect that we can extrapolate to strong coupling where
the spacetime geometry is described by the black string with a longitudinal
wave. The number of BPS states is also independent of all moduli such as
the size of the circle $L$.  Although we assumed that $L$ was large
to facilitate the counting, the final answer holds for all values of $L$
(provided (\ref{slow}) is satisfied).

\subsection{Internal Waves}
\label{ciw}

We now consider the weak coupling limit of a black string with internal waves.
These waves carry
a momentum $r_0^2 \dot f^i/\k$ in the internal $T^4$,
and a momentum $(p + r_0^2 \dot f^2)/\k$ along the string.
Since there are no internal directions in our $1+1$
dimensional description of the effective degrees of freedom for the
black string, the question arises as to how to represent the internal
momentum. There is a natural answer when $Q_5 =1$. In this 
case the fields in the sigma model 
represent fluctuations of the $Q_1$ onebranes (D-strings) inside
the fivebrane.  Thus for each D-string and each internal direction, there
is a 1+1 bosonic scalar field $\chi$ representing the displacement of
the string in that direction.  As a result, the spacetime momentum of
the string in an internal direction must be equal to the
{\it field} momentum
$\dot \chi/ 2 \pi g$, as both generate translations of the
D-string in the internal
direction.
This is not to be confused with the momentum  associated
with translations in the $z$-direction;
this is the longitudinal momentum $T_{uu}$
that we have discussed previously. From (\ref{charge}) and (\ref{defkap})
the internal momentum of the black string is
\be\label{mbs}
{r_0^2 \dot f^i\over \k} = Q_1 {\dot f^i\over 2 \pi g}
\ee
which is indeed just the field momentum of $Q_1$ fields with $f=\chi$!
Since the normalizations of $f$ and $\chi$ are fixed independently, this
agreement is further support for our interpretation of the entropy.
The problem is now to count the number of
BPS states in the $1+1$ dimensional sigma model with field momentum
constrained to be 
$r_0^2\dot f^i/\k$ and longitudinal momentum 
$(p + r_0^2\dot f^2)/\k$.
(The field momentum associated with the fermions is 
unconstrained.)
If we ignore the  constraint on the field momentum,
our previous argument
shows that the number of states with longitudinal momentum $(p +r_0^2
\dot f^2)/\k$ is
much greater than the number with momentum $p/\k$  when $ \dot f^i$ is large.
We will see that when the constraint is included, the number of
states becomes independent of $ \dot f^i$ in agreement with the
Bekenstein-Hawking entropy.

Let us focus on one component of the internal momentum and drop the
index $i$.
We proceed as before by dividing space into intervals $\Delta_a$ on
which $\f$ and $p$ are approximately constant.  We argued above that
each interval  acts like an independent closed string.
For each field $\chi$,
the total field momentum in this interval is given by the zero mode 
\be\label{zzmode}
{1\over 2\pi g} \int_{\Delta_a} g_a \ \p_u \chi = {1\over \sqrt{2g}} 
\alpha_{a,0}.
\ee
Since $\dot f$ is assumed constant on the interval $\Delta_a$,
the contribution of the internal wave to the longitudinal momentum is
\be\label{intwav}
{r_0^2 \over \k} \int_{\Delta_a} g_a \dot f^2 = {Q_1\over 2\pi g} 
\int_{\Delta_a} g_a \dot f^2 = {Q_1 \pi \alpha_{a,0}^2 \over l},
\ee
which is just $Q_1$ times the usual contribution of the zero mode of $f$ to the 
longitudinal momentum.
Thus if there was only one field ($Q_1 = 1$), the internal momentum
$r_0^2 \dot f/\k$ would fix the zero mode, and the momentum left to
distribute to the nonzero modes would be just $p/\k$,
the same as it would be if $\f =0$. 
(Previously, the
zero mode was unconstrained and, as a result,
most of the states
had zero modes which were small compared to the total momentum.)
Since the state
of the zero mode is fixed, it contributes no entropy to the system.
The counting of states thus proceeds just as in
section \ref{clw} and the entropy is
${2 \pi} \sum_a \sqrt{N_a /6 }$ with $ 2\pi N_a/l = p_a l/\k$.

Since there are $Q_1$ different bosonic fields which contribute to the field 
momentum (in a given direction), only the total zero mode
is fixed, not the zero modes of each field.  To compute the entropy,
we proceed as follows. Let us focus on one interval and drop the subscript
`$a$'. Suppose the zero mode of the $A^{th}$ string
is $\alpha_0^A$. Then from (\ref{mbs}) and (\ref{zzmode}) our constraint is 
\be\label{zmode}
\sum_A \alpha_0^A = Q_1{l \dot f\over \pi \sqrt{2g}}.
\ee
Let $k^2= (\pi/l)\sum_A (\alpha_0^A)^2$ be
the contribution of these zero modes to the longitudinal momentum $P$.
Then for a given $k$, the total entropy consists of two contributions:
one from the number of states with longitudinal momentum
$P- k^2$, and the other from the number of ways of 
distributing the zero mode momentum among the strings so that 
their sum is given by (\ref{zmode}) and the sum of their squares is $k^2$.
The latter
contribution can be estimated by considering the allowed volume of phase space.
Our two conditions define a plane and a sphere in a $Q_1$ dimensional space.
These surfaces do not intersect if $k^2 < k_0^2 \equiv Q_1 l \f^2 /2\pi g$. 
For
$k^2 > k_0^2$ they intersect in a sphere of area less than $k^{Q_1}$.
Note that $k=k_0$ represents the point in phase space where
the zero modes are all equal, so the
field momentum is distributed
evenly among all of the strings.   
Since the entropy coming from the nonzero modes
with longitudinal momentum grows much
faster than the entropy coming from the zero modes,
one expects that the total entropy will be maximized when
$k$ takes its minimum value. Setting $k^2 = k_0^2 $ and using the
fact that $k_0^2$ is precisely the contribution of the internal wave
to the longitudinal momentum (\ref{intwav})
we see that the
entropy will be the same as that of a purely longitudinal
momentum distribution $p$ in each interval. Thus the total entropy will be
independent of the internal wave, in agreement
with the Bekenstein-Hawking entropy. 

More precisely, for a general $k$,
the entropy is bounded by
\be
S \le  \sqrt{2\pi Q_1 l(P - k^2)}
+ {Q_1} \ln {k/k_0}
\ee
where the additive constant $-Q_1\ln k_0$ is chosen so that the
zero modes contribute no entropy for $k=k_0$.
Using the fact that $P l >> Q_1$,  the  maximum of $S$
occurs for $k^2  \approx \sqrt {Q_1 P/2\pi l}.$ Recall that
$P = (p + r_0^2\dot f^2)l/\k$.                                                  
If $p$ is not  much larger than $r_0^2\dot f^2$,
the preferred value of $k^2$ is less than the minimum allowed
value $k_0^2 \sim P$. So the entropy is maximized by $k^2=k_0^2$ as expected.
If $p$ is much larger than $ r_0^2 \dot f^2$, the entropy is
maximized for  $k^2> k_0^2$, but in this case the internal waves $\f$ make
only a small correction to the entropy.
The net result is that in all cases the leading order contribution 
to the entropy is  independent of the internal wave 
in agreement with the calculations from the black string.
~
~
\section{Discussion}
\label{disc}

We have shown that the Bekenstein-Hawking entropy of extremal black strings with
traveling waves correctly
counts the number of BPS states in weakly coupled string
theory with  momentum distributions which agree with the black string.
This provides a new
interpretation of the entropy of the black string without traveling
waves: $p=$
constant, $\f_i=0$.
In previous discussions,
the entropy of this black string was reproduced
by counting all BPS states with total momentum $P$. Here we interpret this
solution as representing
only a subset of these states; namely those for which the
distribution of momentum is in fact {\it uniform}, at least on a macroscopic
scale. There is no contradiction 
because the vast majority of states with total momentum $P$ {\it do} in
fact have their momentum evenly distributed.  This can be seen
from the fact that among all momentum distributions $p(u)$ with total momentum
$P$, the entropy (\ref{entropy}) is maximized for constant $p(u)$.

There are additional solutions describing black strings with traveling
waves that we have not yet discussed. For example, we have considered
transverse waves in the internal directions, but not transverse waves
in the asymptotically flat
directions. These are described by $K = -2 \ddot h_i(u) x^i$
cf (\ref{waves}). Black strings with these waves have nonzero
momentum components
in the  four macroscopic directions.
In the Appendix it is shown that these waves are
similar to the internal waves in that they do not affect the horizon
area. This can also be understood by counting string states.
Let us again consider the case $Q_5=1$.
In the weakly coupled D-brane description, there are four bosonic fields
describing oscillations of the fivebrane in the noncompact
directions. Such fields are not usually included in the counting of the
entropy since they add only a small number to the $4Q_1$ fields arising from
the oscillations of the D-strings. However, to reproduce the momentum
in the macroscopic dimensions, one simply fixes the momentum distribution
of this
field to the specified value. As for the internal waves, the field
momentum carried by the zero mode in each small region
will compensate for the extra 
longitudinal momentum so that the number of states is independent of
these transverse waves.

One can also consider solutions where $K$ is quadratic in the transverse
coordinates. If $K$ satisfies Laplace's equation, the matter fields remain
unchanged and the solution describes a
gravitational plane wave superposed on the black string. Alternatively, one
can add plane waves to the matter fields. For example, if one adds an
arbitrary function of $u$ to the dilaton $\phi$, the metric (\ref{inhom})
remains
a solution provided $K$ satisfies
$\p^2 K = 4\ddot \phi$.
The solution for $K$ is again quadratic in the transverse coordinates. Since
this is negligible compared to the $p$ term near the horizon, these plane
waves should 
not affect the horizon area either.
On the one hand, this is consistent with the fact that, 
in the absence of the black string, the plane wave solutions have no
horizon and no gravitational entropy. On the other hand, it is clear that
there is more than one
microstate that can reproduce a given classical wave, so there should be
some entropy associated with the wave. The resolution is simply that the
entropy associated with the classical wave is much smaller than that of
the black string. This can be seen as follows.
In terms of the weakly coupled description, in
addition to the D-brane states, one must now include closed string states
to reproduce the macroscopic plane waves.
Counting such states is much like our counting of states for 
a given internal wave on a  single D-string (having $p(u) = 0$) in section
\ref{ciw}.  In each small region $\Delta_a$ (on which the
macroscopic wave is roughly constant) all of the energy in the 
wave must go to the zero mode on $\Delta_a$; there is no extra (macroscopic)
energy to be distributed randomly among the modes.  As a result, the
entropy of a matter wave is the entropy of `zero thermal energy'.
This depends on the details of one's measuring apparatus, but
is independent of the wave; it is merely an additive constant.  The
same is true of a passing gravitational wave.

For the black string without traveling waves, it was shown in \cite{host}
that the entropy of the slightly nonextremal solution could also be
reproduced by counting string states at weak coupling. The difference
was simply that one must include left moving modes as well as right moving
modes with their respective momenta equal to the left and right moving 
momenta
of the black string. It was argued that if $L$ was sufficiently large,
the energy of the modes carrying the entropy would be small enough so that
interactions remain negligible even when the string coupling is 
increased to form
the horizon. Since the black string with traveling waves corresponds
to just a subset of the right moving modes considered previously,
one can still add left moving modes (with an arbitrary momentum distribution)
in the weak coupling description. What spacetime is produced
when the coupling is now increased? This is puzzling since there does
not appear to be a nonextremal analogue of a black string with traveling
waves. The nonextremal black string does not have a null translational
Killing field, so one expects that any wave present in the spacetime
will either fall into the black hole or disperse to infinity.
A possible explanation is that neither description is really time
independent: The slow process of the nearly extremal
black string absorbing the traveling wave might correspond 
to the slow process
by which the
weak but nonzero interactions between right- and left- movers homogenize
the momentum distribution in the weakly coupled string.

In \cite{ght} it was shown that the extremal black string (\ref{simple})
with no momentum ($p=0$)
and no traveling waves has an unusual global structure. There is
an event horizon at $r=0$ but no singularity inside. In fact the region
inside the horizon turns out to be  another asymptotically flat spacetime
which is isometric to the region outside.\footnote{When the momentum is
zero, the horizon area of the compactified spacetime vanishes and there is a
conical singularity on the horizon resulting from the periodic identification of
$z=(v-u)/2$.} The
addition of any momentum along the string changes this behavior. It is
shown in the Appendix that the black string with nonzero momentum
has a singularity inside the horizon. This should be expected since
the dimensional reduction of one special case of this solution
yields the five dimensional
Reissner-Nordstr\"om metric, which is certainly not symmetric
across the horizon.

The four dimensional Reissner-Nordstr\"om metric has been 
shown to arise from the dimensional reduction of a six dimensional
black membrane which is similar to the  black string discussed here \cite{cvts}.
In particular, near the horizon the solution again reduces to the
product of three dimensional anti-de Sitter space and $S^3\times T^4$. The
entropy of the translationally invariant  solution (corresponding to
a uniform momentum density) has been reproduced by counting states
at weak string coupling. One can add traveling waves to this solution
and we expect that the entropy will again correctly count the number of
string states. This is currently under investigation.

\acknowledgements
It is a pleasure to thank J. Louko, J. Harvey, and
J. Polchinski for useful discussions. This
work was supported in part by NSF grant PHY95-07065.

\appendix
\section{Details of the exact metric}

As mentioned in section \ref{metric}, our study of the leading order
behavior (\ref{hor}) of the black string metric is not really
sufficient to show that the horizon is a smooth surface and, as a
result, it is also insufficient to show that the 
area of the horizon is
$2\pi^2 r_0^4 L {\sf V}
\overline{\sigma}$.  The point is that while a study of 
(\ref{hor}) shows that the leading order divergences in (\ref{inhom})
are purely coordinate effects, any sub-leading divergences 
still remain to be addressed.  In this appendix, we consider
the exact metric, showing that the horizon is smooth when the momentum
distribution is uniform along the string, that it is at least $C^0$
for a general metric of the form (\ref{inhom}) with $K$ given by
(\ref{waves}), and that 
in both cases the area of the horizon is given by
$2\pi^2 r_0^4 L {\sf V} \overline{\sigma}$.  In addition, we comment
on the global structure of the homogeneous black string and we extend our
treatment to include `transverse waves' in the macroscopic
directions (for which $K = - 2 \ddot{h}_i(u) x^i$), showing that the results
stated above hold for these waves as well.  
{}From \cite{DGHW}, the metric with such transverse waves can be placed in 
an asymptotically flat form by a coordinate transformation similar to
(\ref{afct}).  

Our discussion will parallel that of section \ref{metric}.  We
begin by writing the exact metric as
\begin{eqnarray}
\label{exact}
ds^2 &=& r_0^2 \biggl\{- R^{-2}[dud\hat{v} 
-2 \dot{f}_i dy^i du -2 \dot{h}_i dx^i du] \cr
&+& {{p(u)} \over {r_0^4}} du^2
+ R^{-2} Z^{-3} dR^2 + Z^{-1} d^2\Omega_3 
\biggr\} + dy_i dy^i
\end{eqnarray}
in terms of the coordinates
$u$, $\hat{v} = v +2 \dot{f}_i y^i +2 \dot{h}_i x^i + r_0^{-2} \int^u p du $,
and $R = r_0 \sqrt{{r^2 + r_0^2} \over {r^2}}$.  Here, 
$Z(R) = {{R^2 - r_0^2} \over {R^2}} = 1 - {{r_0^2} \over {R^2}}$ 
takes the value $1$ on the horizon; the departure of $Z$ from this
value represents the terms neglected in the leading-order expression
(\ref{hor}).  Note the slight change from section \ref{long} in the
definition of $R$.

We now introduce new coordinates $U = F(u)$, $V = \hat{v} - R^2
\sigma(u) - Y(u)$, and
$W = G(u)/R$ where, as before, $\dot F = G^{-2}$, $
\sigma(u) = G^{-1}\dot G$,
and $\sigma^2 + \dot \sigma = r^{-4}_0p$; $Y(u)$ will be specified
below.  Recall that the horizon $u=\infty$ lies at a finite value
of $U$; we again choose this value to be $U=0$.  The metric now becomes
\begin{eqnarray}
\label{exactii}
ds^2 &=&  r_0^2 \biggl\{ -W^2 dUdV +2 W^2 \dot{f}_i dy^i dU  +2  W^2
\dot{h}_i dx^i dU - G^2 [\sigma^2(1-Z^{-3})G^2 + W^2 \dot{Y}]
dU^2 \cr
&+& dUdW[2W^{-1}\sigma G^2 (1 - Z^{-3})]
+ W^{-2}Z^{-3} dW^2 + Z^{-1} d\Omega^2_3 \biggr\} + dy_i dy^i.
\end{eqnarray}
Since $G$ is roughly exponential in $u$, we see that the
$dU^2$ and $dUdW$ terms have
several potential divergences at the horizon $u = 
\infty$.  Since $Z = 1 - W^2 r_0^2 G^{-2}$, $g_{UW}$
is of order $G^0$ while $g_{UU}$ is of order $G^2$.  However, 
by choosing $Y(u)$ to satisfy $dY/du = 3r_0^2 \sigma^2 $ we can
arrange that $g_{UU}$ is also of order $G^0$.  For the case where
$\dot{f}_i = 0$ and
$p$ is independent of $u$, $\sigma = r_0^{-2} \sqrt{p}$ is a
constant and the metric
(\ref{exactii}) is smooth (and even analytic) at the horizon; the
area of the horizon may be computed as in section \ref{metric}
and agrees with the result stated there.  

We can also see that (for $\dot{f}_i = 0$, $p = constant$)
the global structure for $p \neq 0$ is much different from the 
$p=0$ case,  
which has no singularity anywhere in the uncompactified spacetime
\cite{ght}.
In fact, it is symmetric across the horizon; passing through the
horizon leads to an asymptotically flat region which is just like the
original one.  This is {\it not} the case for the solution
(\ref{exactii}).  Note that the
radius of the 3-spheres is $r_0 Z^{-1/2}$.  Now, when 
$p$ is uniform, $G= e^{\sqrt{p}u/r_0^2}$ and $U = - {{r_0^2} \over
{2 \sqrt{p} G^2}}$.  As a result, $Z = 1 + 2 \sqrt{p} W^2 U$ is less
than one outside the horizon ($U<0$), but greater  than one
inside ($U>0$).  Thus, the 3-spheres continue to shrink as the horizon
is crossed and a singularity forms when they reach
zero size.

Returning to the general case, suppose that 
$p$ is periodic in $u$ but not constant.  In this case $p$, 
$\dot{f}$, and $\dot{h}$ (and
therefore $\sigma$) oscillate infinitely many times before reaching
the horizon at $U=0$; as a result, the expression (\ref{exactii})
is not even continuous at the horizon.  Another source of concern
is the fact that 
\begin{eqnarray}
dx^i &=& {{x^i} \over r} dr + r \ d(angles)^i \cr
&=& - \left( {{x^i} \over r} \right) {{r_0^2} \over {Z^{3/2}}}
\{ W\dot{G} dU - G^{-1} dW \} + r \ d(angles)^i 
\end{eqnarray}
for which the $dU$ term diverges for $U \rightarrow 0$.
These are again  coordinate
effects.  To see that this is so, it is simplest to work with the
leading order (${\cal O} (G^0)$) behavior of (\ref{exactii}).  
Since $G$ diverges at the horizon, any higher order terms vanish
at $U=0$ and so are already continuous there\footnote{Moreover, they
remain continuous under the coordinate transformations performed below.}.
We therefore write the metric
(\ref{exactii}) in the form
\begin{eqnarray}
\label{pert}
ds^2 &=& r_0^2 \biggl\{ -W^2 dUdV 
+2  W^2 \dot{f}_i dy^i dU -2 W^3 \left( {{x^i} \over r} \right)
\dot{h}_i r_0^2 \dot{G} dU^2 \cr
&+& 6 W^4 r_0^4 \sigma^2 dU^2 - 6 W r_0^2 \sigma
dUdW + W^{-2} dW^2 + d\Omega^2_3
\biggr\} + dy_i dy^i + ds_0^2
\end{eqnarray}
where $ds_0^2$ contains only higher order $C^0$ terms. The
metric can now be written in $C^0$ form by performing
a final coordinate transformation.  It is simplest to break this
into a series of steps.  For convenience, let $\beta_i =
\int^U \dot{f}_i dU$.  Transforming to coordinates
$(U,\tilde{V},q,\theta,\tilde{y})$ where
\begin{eqnarray}
\tilde{V} &=& V +2 W \left( {{x^i} \over r} \right) r_0^2 \int^U 
\dot{h}_i \dot{G} dU' \cr
q &=&- ( {1 \over 2} W^{-2} + 3 r_0^2 \int^U \sigma dU ) \cr 
\tilde{y}^i &=& y^i + r_0^2 W^2 \beta^i
\end{eqnarray}
results in the metric
\begin{eqnarray}
ds^2 &=& ds_0^2 - r_0^2 W^2 dU d \tilde{V} - r_0^4 W^4 (\dot{f}^2
+ 3 r_0^2 \sigma^2) dU^2 + r_0^2 d\Omega^2_3 + \cr &+& r_0^2 W^4 dq^2 +
( d\tilde{y}^i - 2 r_0^2 W^4 \beta^i dq - 6 r_0^4 W^4 \beta^i \sigma
dU)^2.
\end{eqnarray}
Since $W$ is a $C^0$ function of the new coordinates, a
further transformation to coordinates $(U,\nu,q,\theta,z)$ where
\begin{eqnarray}
\label{final}
\nu &=& \tilde{V} + \int^U r_0^2 W^2 (\dot{f}^2  + 3 r_0^2 \sigma^2) dU'
\cr
z^i &=& \tilde{y}^i - \int^U 6 r_0^4 W^4 \sigma \beta^i dU'
\end{eqnarray}
places the metric in $C^0$ form.  In (\ref{final}),
the integrals are performed at fixed $q$ (not fixed
$W$).   
Again the final form of the metric is not particularly enlightening.
Whether or not the metric is in fact {\it smooth} at the horizon is a
question which deserves further study.

Note that we have not replaced the coordinate $U$ in obtaining the
$C^0$ form.  As a result, 
the horizon does indeed lie at $U=0$ and the metric on the horizon
can be read directly from (\ref{exactii}); it is just $r_0^2 W^{-2} dW^2
+ r_0^2 d\Omega^2_3 + dy_idy^i$.
It follows that the horizon is homogeneous and that it
has 
area $2\pi^2 r_0^4 L\V \overline{\sigma}$ as claimed in section \ref{metric}.

\end{document}